# Cooperativity of short-time dynamics revisited

D. Fragiadakis and C.M. Roland

Naval Research Laboratory, Chemistry Division, Washington DC  20375-53421

*(January 18, 2018)*

ABSTRACT. Using molecular dynamics simulations we examine the system size dependence of the fast dynamics in two model glass forming liquids, one of them a Lennard-Jones mixture for which cooperative fast relaxation has been reported. We find no indication of a temperature-dependent dynamic length scale characterizing these fast dynamics; the size effects in the short time range are temperature independent, and the consequence of cutting off of long wavelength acoustic modes. In a molecular liquid exhibiting a clear Johari-Goldstein (JG) relaxation, significant size effects are again present both for the vibrational motion and long-time α relaxation (only the latter having a significant temperature dependence), but absent for the JG relaxation.

Glass-forming liquids exhibit complex dynamics, with motions taking place at multiple timescales. Displacements and reorientations that alleviate perturbations are referred to as relaxations. One approach to understanding glass formation is to characterize the relaxation behavior at times prior to structural ($\alpha$-) relaxation, and assess any relationship between the structural and fast dynamics. Studies carried out along these lines have concluded that a connection indeed exists, manifested ultimately in the fast dynamics "sensing" the glass transition [1,2,3,4,5,6]. This approach requires identification of relevant secondary relaxations, distinct, for example, from trivial pendant group motion. Of the various high frequency processes, the most important for understanding glass formation is the secondary $\beta$ relaxation. It is most commonly detected by dielectric spectroscopy, since even small fluctuations in molecular orientation can yield a significant change in local polarization. Secondary dynamics involving motion of the entire molecule are referred to as the Johari-Goldstein (JG) relaxation. Unlike other secondary processes, the JG relaxation is found universally in glass-forming materials, including liquids, polymers, ionic conductors, and metallic glasses. It has properties that correlate with those of the $\alpha$ relaxation [7], which leads to the idea that the JG process serves as the precursor of the glass transition; that is, the short-time JG dynamics evolves into structural relaxation transpiring at much longer times (>10 s) [8].

Since it is well established that the α relaxation is cooperative, governed by dynamic heterogeneity and intermolecular correlations that increase on cooling [9], the question arises whether the JG regime is cooperative.

Intuitively the expectation is that going from vibrational caging at very short timescales to the JG process (observed dielectrically in the kHz – MHz frequency range), then ultimately to the α relaxation, will be associated with increasing length scales; therefore, the JG dynamics would be at most weakly cooperative. (The coupling model of Ngai [8] makes this assumption explicitly in connecting the JG relaxation time to the non-cooperative time constant of the model.) The JG relaxation is dynamically heterogeneous, but dynamic heterogeneity does not imply cooperativity [10]. Johari and Goldstein originally described the process as involving rearrangements of "at least one, but probably several molecules" [11]. The high activation energy and activation entropy of the JG process in the glassy state have been interpreted as indicating some degree of cooperativity [12]. To distinguish this from the strong intermolecular cooperativity exhibited by the α relaxation, the β process has been called "locally coordinative" [13]. Consistent with this idea, it has been found that the potential barriers for the JG relaxation overlap, although they may be somewhat lower than the barrier heights for the α relaxation [14]. An interpretation of the JG process as rearrangements of string-like clusters of molecules implies cooperativity [15], and some degree of intermolecular cooperativity is inferred from NMR measurements of binary mixtures [16] and organic phospate glasses [17]. In molecular dynamics (MD) simulations of asymmetric dumbbell-shaped molecules, a weak maximum in the dynamic susceptibility at the timescale of the JG relaxation indicates some degree of dynamic correlation, but it is very weak compared to the correlation of the α relaxation, and its intensity does not increase on cooling (in fact in the glassy state dynamic correlations on the JG timescale decrease on cooling) [18].

Recently MD simulations have been carried out focusing on the short-time relaxation dynamics in Lennard-Jones and other model systems [19]. Following Stein and Andersen [20], in these studies the β relaxation is associated with the minimum in the derivative of the logarithm of the mean square displacement (MSD) with respect to the logarithm of time. A length scale is determined from finite-size scaling, which is found to have the same temperature dependence as the length scale describing spatial heterogeneity of the α relaxation (both increasing on cooling). The interpretation of these results has been that this β relaxation is cooperative, with likely the same origin as the α relaxation evident at the long-time end of the MSD plateau [19]. The systems studied previously [19] do not show an explicit signature of the JG relaxation as an increase in the MSD or a step in the intermediate scattering function.

The process ostensibly transpiring with the MSD plateau is not the JG relaxation, as evident from behavior quite different from that of the latter [8,8,11,12,13,14,15,16,17]. It is not obvious what underlies this putative secondary relaxation, given the absence of dynamics in the MSD plateau.

In this short paper we examine in detail which features of the dynamics in the binary LJ system manifest as an inflection point in the MSD. We identify two such features, distinct from each other, providing an alternative explanation for the results in ref. [19]. In addition, we employ MD simulations of diatomic molecules to study the secondary dynamics. For diatomic molecules, unlike the L-J particles of ref. [19], the orientational correlation function shows a distinct β process, manifested as a peak in the frequency domain [21,22]. The correspondence to experimental dielectric spectra of substances exhibiting JG dispersions is apparent. This unambiguous detection of the JG relaxation in simulations enables observation of the features that define the JG process in real materials, such as merging with the α peak at high temperature, changes in the strength of the α relaxation at $T_g$, and significant sensitivity to density, pressure, and physical aging [22]. These properties distinguish the JG relaxation from trivial secondary dynamics that involve only intramolecular degrees of freedom [7,23] and from the fast β process of mode coupling theory [24,25,26].

Simulations were carried out using the RUMD simulation software [27]. All simulations were performed in the NVT ensemble with a Nose-Hoover thermostat [28]. Two systems were simulated: (a) the three-dimensional Kob-Andersen binary Lennard-Jones mixture (KABLJ) [29], with system sizes ranging from N=100 to N=65,000 particles at density $\rho = 1.2$ and temperatures in the range T=0.1 to T=1.0. For temperatures above T=0.45, the system was well equilibrated using an NVT run several times longer than the α relaxation time. For T<0.45, the system was equilibrated for t=500,000, shorter than the α relaxation time; therefore, these represent out-of-equilibrium glass. (b) Asymmetric diatomic molecules (AD) [21], in which each molecule is composed of two Lennard-Jones particles with size parameters 1.0 and 0.625 connected by a rigid bond. The bond length was maintained constant at l=0.45 using a constraint force algorithm [30]. System sizes from N=50 to N=5000 molecules were simulated at temperatures from T=0.3 to T=0.6 at a number density of $\rho = 1.2$. A cutoff of $r_{cut}$=2.5 units was used for the Lennard-Jones interactions.

Figure 1 shows the system size dependence of the MSD and its derivative $d\log\langle\Delta r^2\rangle/d\log t$ for the KABLJ system at two temperatures. With decreasing system size, the height of the MSD plateau decreases, with an "overshoot" developing at the beginning of the plateau. The overshoot is a finite size effect caused by the cutting off of long-wavelength acoustic sound modes due to the finite size of the simulation box [31]. More prominent in 2D than 3D simulations, the overshoot is manifested as a "bump" in the intermediate scattering function seen in the Kob-Andersen mixture [28], as well as other systems such as silica [32] and OTP [33]. The time of the overshoot, $t_o$, is temperature-independent and scales linearly with the length of the simulation box. A connection of the overshoot to the Boson peak has been suggested [31,32], and it may have the same origin as the short-time peak in the 4-point dynamic susceptibility, the time scale of which also increases linearly with system size [19]. At T=0.45 the minimum in the derivative (denoted by triangles in Fig. 1) corresponds to this overshoot, located near the beginning of the MSD plateau. At T=0.55, the plateau ends at shorter times, and for the larger system sizes (N>2000) the minimum in the derivative no longer coincides with the overshoot; rather, the former moves towards the middle of the plateau at $t_{mid}$ (> $t_o$). These observations make evident that the inflection in the MSD and its temperature dependence are not properties of a putative secondary relaxation; rather, they reflect unrelated, piecemeal effects, the relative contribution of each depending on state point and MSD box size.

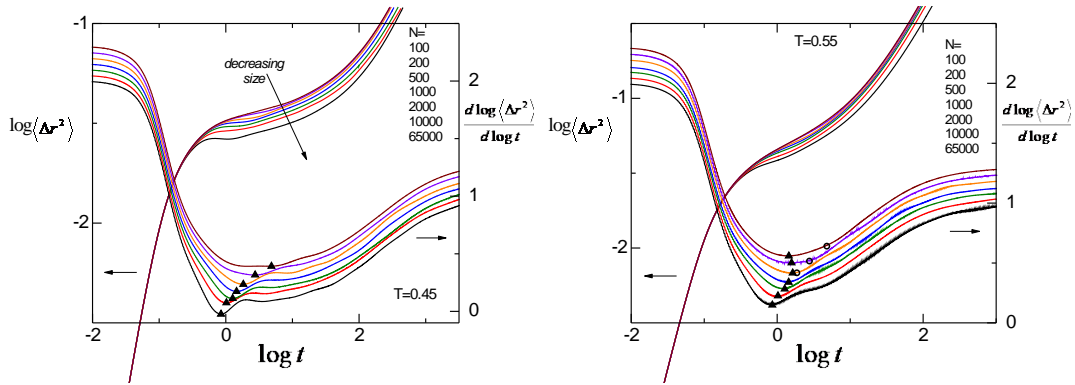

FIG. 1. (left) System size dependence at T=0.45 of MSD and the derivative of log MSD with respect to log t (the latter shifted vertically for clarity), for the KABLJ system. The MSD begins with slope=2 (ballistic), develops a plateau, and eventually increases with slope=1 (diffusion). Triangles denote the minimum in $d\log\langle\Delta r_2\rangle/d\log t$. (right) Same data at T=0.55, higher temperature causing the plateau to terminate at shorter times. The overshoot determines the temperature of the minimum for N<2000, but at N>2000 the overshoot (circles) is barely visible, with a change of slope in the derivative; the minimum falls in the middle of the plateau.

Figure 2 shows the MSD and $d\log\langle\Delta r^2\rangle/d\log t$ at a series of temperatures for different system sizes (number of particles). At high temperatures the derivative minimum occurs at the middle of the plateau, $t_{mid}$. As temperature decreases and $t_{mid}$ becomes significantly longer than the overshoot time, the minimum is located at $t_o$. Thus, the inflection in the MSD shifts from the plateau midpoint at higher temperatures to the vicinity of the overshoot at lower T. These results are collected in Figure 3, displaying the time of the minimum versus system size. Note that for temperatures below T=0.45, the effect of system size on the fast dynamics is qualitatively the same, despite the fact that at lower temperatures the system is out of equilibrium. The temperature dependence of $t_{min}$ has no specific physical interpretation, but is merely a time intermediate between the vibrational timescale defining the beginning of the plateau and the α relaxation time defining its end. At lower temperatures the minimum follows $t_0$, seguing to a constant $t_{min}=t_{mid}$ at higher T. The parameter $t_{mid}$ very weakly increases with system size, similar to the behavior for the α relaxation.

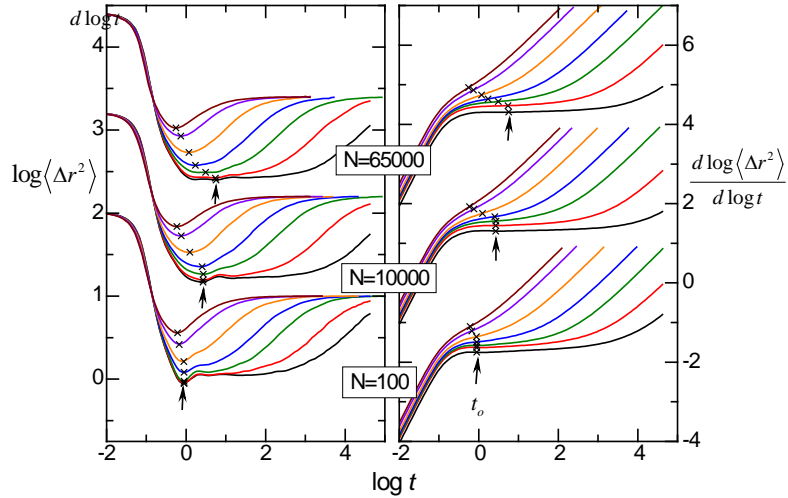

FIG. 2. Same data as Fig. 1 at temperatures (left to right) T= 1.0, 0.8, 0.6, 0.5, 0.45, 0.4, 0.3. The curves are shifted vertically for clarity. At high temperatures (higher plateaus) the derivative minimum $t_{min}(T)$, denoted by crosses, occurs at the middle of the plateau, at a temperature-dependent time $t_{mid}(T)$. As temperature decreases and $t_{min}(T)$ becomes significantly longer than the overshoot time $t_o$ (which depends on N but is temperature independent), the minimum occurs at $t_o$ (indicated by arrows).

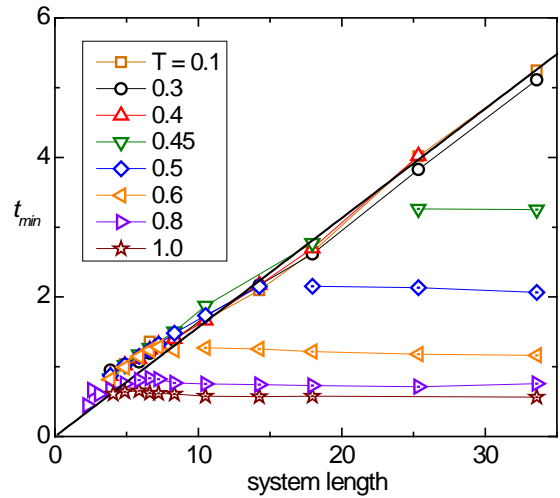

Figure 3: Time of the minimum of $d\log\langle\Delta r_2\rangle/d\log t$ for the KABLJ system versus linear system size (side of the cubic simulation box). Dotted symbols: minimum at $t_{mid}$; open symbols: minimum at $t_o$. The solid line is a linear fit to the $t_{mid}$ data, passing through the origin. (Note that at temperatures T<0.45 where the system is out of equilibrium, the system size dependence of the minimum does not change.)

We also applied finite-size scaling to MD simulations of a diatomic molecule (AD system) that exhibits a genuine JG relaxation (Figure 4). In addition to the MSD and its derivative, we show the imaginary part of the first order rotational susceptibility as a function of angular frequency

$$\chi(\omega) = 1 + i\omega \int_0^\infty dt\, e^{i\omega t} C_1(t)$$

where $C_1(t) = \langle \cos\theta(t) \rangle$ is the first order rotational susceptibility, and θ the angular change in the molecular axis. In the susceptibility plot three peaks are observed, corresponding (from high to low frequencies) to vibrational motions, the JG relaxation, and the α relaxation. The JG relaxation manifests in the MSD as a broad step in the mean square displacement within the plateau region, and a second broad minimum in $d\log\langle\Delta r^2\rangle / d\log t$. On reducing the system size, the α relaxation time increases, opposite to the behavior of the vibrational peak. The latter moves to higher frequency with a decreasing intensity on the low frequency side of the peak. This is due to truncation of the long-wavelength acoustic modes by the finite system size. However, the JG process shows no change, even for a system size as small as 50 molecules; i.e., a box having dimensions less than 4 molecules. (Note further reducing the box size would intrude on $r_{cut}$ used for the LJ interactions and thus lead to artifacts.) We observe a similar invariance to system size for the JG dynamics at lower temperature.

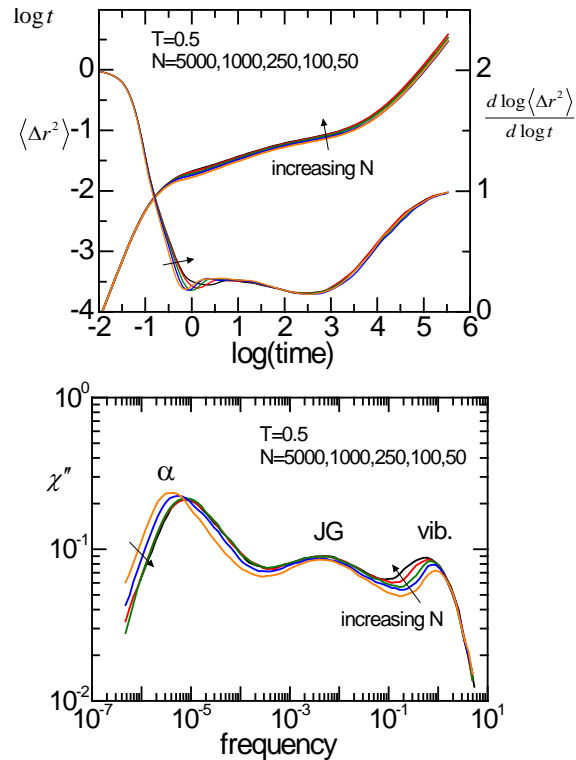

Figure 4: System size dependence at T=0.5 of (top) MSD and the derivative of log MSD with respect to log t and (bottom) imaginary part of the 1st order rotational susceptibility, for the AD system. From high frequencies to low, three peaks corresponding to vibrations, JG, and α relaxation are observed.

In summary, the present results show that the secondary dynamics observed in MD simulations depends on the system studied because of the requirement to have resolved processes that fall within an accessible window of time. Interpretation of the MSD inflection must account for two disparate effects in order to yield reliable information, but whether it is not obvious how the mechanism of the inflection can be connected to an actual relaxation process. The time at which an inflection occurs in a log-log plot of the MSD versus time does not define a physically meaningful relaxation time, nor does analysis of the inflection yield information about the dynamic length scale of short-time motions or their temperature dependence. The inflection corresponds to either to an overshoot in the MSD or the midpoint of the plateau, depending on the temperature and system size. At lower temperatures the inflection follows the overshoot caused by finite-size scaling and the consequent cutting off of long wavelength modes. These changes in the location of the inflection are unrelated to the length scale of any supposed $\beta$ dynamics. On the other hand, at higher T and/or larger system size, for which the overshoot becomes negligible, the inflection falls in the middle of the plateau between the vibrations and the $\alpha$ process; consequently, its T-dependence trivially follows that of the $\alpha$ relaxation. Unlike MD simulations of LJ particles, diatomic molecules exhibit an explicit signature of the JG relaxation. However, there is no indication of a dynamic length scale revealed by finite size scaling, presumably because the smallest feasible box size (four molecular diameters) is too large.

This work was supported by the Office of Naval Research.